\documentclass[sigconf,natbib=true,screen=true]{acmart}

\setcopyright{none}
\copyrightyear{2024}
\acmYear{2024}
\acmDOI{}
\acmISBN{}

\acmConference[CONSEQUENCES]{CONSEQUENCES Workshop at RecSys '24}{October 14--18}{Bari, Italy}
\acmBooktitle{CONSEQUENCES Workshop at RecSys '24, October 14--18, 2024, Bari, Italy}

\author{Morris de Haan}
\orcid{0009-0005-3957-4073}
\affiliation{%
  \institution{University of Amsterdam}
  \city{Amsterdam}
  \country{The Netherlands}
}
\email{morrisdh18@gmail.com}

\author{Philipp Hager}
\orcid{0000-0001-5696-9732}
\affiliation{%
    \institution{University of Amsterdam}
    \city{Amsterdam}
    \country{The Netherlands}
}
\email{p.k.hager@uva.nl}

\usepackage[inline]{enumitem}

\begin{document}

\title{Understanding the Effects of the Baidu-ULTR\\Logging Policy on Two-Tower Models}


\maketitle

\section{Introduction}

The two-tower model is a popular tool for unbiased learning to rank (ULTR) in industry applications due to its simplicity and reported effectiveness~\cite{TwoTowers,RevisitingTwoTowers,DropoutMethod}. The model employs two neural networks called towers. One tower predicts unbiased relevance using query-document features, while the other predicts examination behavior based on bias-related features such as position or device type. By jointly training both towers on click data, the goal is to yield a relevance tower that is free of biases and more effective at ranking.

However, recent work argues that this factorization of bias and relevance assumes independence between the two factors, which is often unrealistic in practice~\cite{DropoutMethod,UPE}, and may degrade the performance of two-tower models. Specifically, when the previous ranker used to collect clicks (the logging policy) ranks relevant documents higher, a correlation between relevance and position bias is introduced, violating the independence assumption. The logging policy acts as a \textit{confounder}~\cite{DropoutMethod,UPE}.

\citet{DropoutMethod} first describe the problem as a case of shortcut learning: Position in many production systems is a strong indicator of relevance, which enables two-tower models to predict relevance based on bias features rather than learning a relevance signal from query and document features. The authors propose two solutions: The first uses a gradient reversal approach from the domain adaptation literature, and the second discourages shortcuts by applying dropout~\cite{Dropout} to the bias tower. In recent work, \citet{UPE} performed a causal analysis of the problem and proposed a backdoor adjustment to mitigate the confounding problem.

Recently, Baidu released a large-scale real-world click dataset for ULTR~\cite{Baidu}, which \citet{Supervisor} used to conduct a reproducibility study of the ULTR field. Their additive two-tower model only achieves marginal ranking improvements over a naive ranker without any bias correction. As a potential explanation, the authors mention logging policy confounding. As previous work on confounding in two-tower models was primarily evaluated in simulation, we revisit the confounding problem on Baidu-ULTR.

We first evaluate if the conditions for the confounding problem are met on Baidu-ULTR by estimating the strength of the logging policy. If the logging policy performs better than random, we expect confounding to impact a two-tower model negatively. Next, we evaluate if two previously proposed methods addressing logging policy confounding improve the ranking performance of two-tower models. Our work answers the following questions:

\begin{enumerate}[nosep, leftmargin=*, label=\textbf{(RQ\arabic*)}]
    \item \textit{What is the ranking performance of the policy that collected the Baidu-ULTR dataset?}
    \item \textit{What is the effect of logging policy-induced confounding on additive two-tower models on Baidu-ULTR?}
\end{enumerate}

\noindent Our code and data for reproducing this work are available at:\\\url{https://github.com/morrisdehaan/two-towers-confounding}.

\vspace*{-0.5em}
\section{Background}
In the following, we introduce two-tower models and two solutions to the confounding problem. We refer to~\cite[Section 2.2]{DropoutMethod} for a detailed description of logging policy confounding.

\vspace*{-0.5em}
\subsection{Two-tower architecture}
This paper assumes the popular additive form of the two-tower model~\cite{TwoTowers,RevisitingTwoTowers,DropoutMethod}, which we define as:
\begin{equation}
    P(C = 1 \mid q, d, k)=\sigma \left( \text{relevance}(x_{q,d}) + \text{bias}(k) \right)
\end{equation}
where $C=1$ denotes a click, $x_{q,d}$ are relevance-related features for document $d$ and query $q$ (e.g., bm25, tf-idf, etc.) and $k$ are bias-related features. In this work, the only considered bias feature is document position. The relevance and bias towers are represented by feed forward neural networks predicting logits which are added and transformed to a click probability using the sigmoid function.

\vspace*{-0.5em}
\subsection{Solutions to logging policy confounding}
The first proposed solution by \citet{DropoutMethod} adds a dropout layer to the output of the bias tower in order to disincentivize learning relevance information through shortcuts. Formally, this yields:
\begin{equation}
    P(C = 1 \mid q, d, k) = \sigma \left( \text{relevance}(x_{q,d}) + \text{dropout} \left(\text{bias}(k), \tau \right) \right)
\end{equation}
with the dropout probability $\tau$, which is a tunable hyperparameter.

In another recent work, \citet{UPE} describe that logging policy confounding leads to \textit{overestimating} examination probabilities in two-tower models. The authors suggest a training procedure based on the backdoor adjustment from causal inference~\cite{backdoor} to \textit{debias} a given set of (biased) examination probabilities:
\begin{enumerate*}[label=(\roman*)]
    \item The logging policy is approximated using query-document features to predict logged positions (similar to our RQ1).
    \item Next, a frozen intermediary embedding from step~(i) is combined with the output of a newly initialized bias tower to jointly approximate the given (biased) examination probabilities.
    \item The resulting embedding is supposed to be immune to the confounding problem and is used as the bias tower in a two-tower setup.
\end{enumerate*}
We refer you to \cite[Section 4.2]{UPE} for more details.

\vspace*{-0.5em}
\subsection{The Baidu-ULTR dataset}
This research uses the preprocessed subset of Baidu-ULTR released by~\citet{Supervisor}, which includes $768$ dimensional BERT-based~\cite{BERT} query-document embeddings and classical LTR features (e.g., bm25, tf-idf, etc.)\footnote{\url{https://huggingface.co/datasets/philipphager/baidu-ultr_uva-mlm-ctr}}. The dataset is split into four partitions of click data with $4.8$m query-document pairs each. We evaluate the final ranking performance on the official Baidu test set containing $380$k query-document pairs with expert relevance annotations.

\vspace*{-0.5em}
\section{RQ 1: Estimating the Logging Policy}
To evaluate if the problem of logging policy confounding exists in Baidu-ULTR, we first need to estimate the ranker that collected the dataset (RQ1). Therefore, we train a LambdaMART model~\cite{LambdaMARTOverview} to reproduce the logged document positions provided in the click dataset based on query-document features. We evaluate how \emph{accurately} we can predict the logged document order in terms of nDCG@10~\cite{ndcg} on one test partition of the click dataset. A model perfectly approximating the top-10 results of the Baidu logging policy would achieve an nDCG@10 of 1.0.

Second, we evaluate the \emph{strength} of the estimated logging policy in terms of ranking performance on the test set of expert annotations and compare it to a random ranker, which serves as a lower bound. Hyperparameter tuning and training details are listed in Appendix~\ref{appendix:logging-policy}.
\begin{table}[h!]
    \vspace{-0.5em}
    \centering
    \caption{nDCG@10 performance of our estimated Baidu-ULTR logging policy and a random baseline with a $95$\% CI.}
    \vspace{-0.3cm}
    \begin{tabular}{lcc}
        \toprule
        Method / nDCG@10 & Accuracy & Strength \\ 
        \midrule
        LambdaMART & $\boldsymbol{0.933\pm4.4\cdot10^{-4}}$ & $\boldsymbol{0.353\pm6.1^{-3}}$\\
        Random & $0.705\pm6.5\cdot10^{-4}$ & $0.317\pm5.7^{-3}$\\ 
        \bottomrule
    \end{tabular}
    \label{table:rq1}
    \vspace{-0.5em}
\end{table}
Table~\ref{table:rq1} shows that LambdaMART achieves near-perfect nDCG when predicting the logged ranking, reasonably approximating the ranker that collected Baidu-ULTR. Additionally, the model achieves much higher ranking performance than the random ranker on expert annotations, indicating that the Baidu logging policy likely introduces a correlation between position bias and relevance features. This, in turn, suggests a confounding problem in Baidu-ULTR, answering RQ1.

\vspace*{-0.5em}
\section{RQ 2: Two-Tower Confounding}
After establishing the likely presence of the confounding problem in Baidu-ULTR, we evaluate if the proposed dropout and backdoor adjustment methods improve the performance of a two-tower model. Additionally, we include performance-wise lower bounds to assess the effectiveness of the two-tower models:
\begin{description}
    \item[Random:] Generates rankings by random shuffling, which any model learning anything substantive should outperform.
    \item[Naive:] A relevance tower without the bias tower. If the bias tower is beneficial, two towers should outperform this naive model.
    \item[Expert:] A LambdaMART trained on expert annotations using k-fold cross-validation~\cite{kfold-cross-validation}. A common assumption in ULTR is that models learning from clicks outperform (or match) such a model given enough time and data.
\end{description}
\begin{table}[h]
    \vspace{-0.5em}
    \centering
    \caption{Ranking performance on Baidu-ULTR with a $95$\% CI. We include DCG@10 for comparability with \cite{Supervisor, Baidu}.}
    \vspace{-0.3cm}
    \begin{tabular}{lcc}
        \toprule
        Method & $nDCG@10$ & $DCG@10$ \\ 
        \midrule
        Random & $0.317\pm5.7^{-3}$ & $6.67\pm1.5^{-1}$\\ 
        Naive & $0.366\pm6.3^{-3}$ & $7.46\pm1.6^{-1}$\\ 
        LambdaMART (expert labels) & $\boldsymbol{0.539\pm7.0^{-3}}$ & $\boldsymbol{10.95\pm2.0^{-1}}$\\
        LambdaMART (logging policy) & $0.353\pm6.1^{-3}$ & $7.33\pm1.6^{-1}$\\ \midrule
        \textbf{Two Tower} & $\boldsymbol{0.375\pm6.3^{-3}}$ & $\boldsymbol{7.62\pm1.6^{-1}}$\\ 
        Two Tower (dropout) & $0.369\pm6.3^{-3}$ & $7.49\pm1.6^{-1}$\\ 
        Two Tower (backdoor adjust.) & $0.366\pm6.3^{-3}$ & $7.43\pm1.6^{-1}$\\ 
        \bottomrule
    \end{tabular}
    \label{table:rq2}
    \vspace*{-0.5em}
\end{table}
Details about hyperparameter tuning and training are listed in Appendix~\ref{appendix:two-tower-confounding}. Table~\ref{table:rq2} compares the ranking performance on the expert annotated Baidu-ULTR test set. First, we note that all models outperform the random baseline. Second, we also note that all two-tower models outperform the estimated logging policy. Third, we observe a small but significant improvement from the standard two-tower model over the non-debiasing naive model, echoing findings by \citet{Supervisor}. Surprisingly, however, we find that both confounding correction methods did not improve the ranking performance of the two-tower model but even decreased the performance in our experiment. Both models seemingly converge to a performance similar to the naive model. Lastly, and most strikingly, we observe that all models trained on click data are vastly outperformed by the expert model trained on a much smaller dataset of expert annotations (with cross-validation). Answering RQ2, our results show no evidence for logging policy-induced confounding of the two-tower model on the Baidu-ULTR dataset.

\vspace*{-0.5em}
\section{Discussion}
While the theoretical conditions for the confounding problem are met because the Baidu logging policy likely correlates position bias and relevance, our results indicate no empirical evidence for the confounding problem. Namely, the standard two-tower model outperforms the proposed solutions and shows no signs of collapse in performance. Even when analyzing the results by logging policy strength (Appendix~\ref{appendix:logging-policy-buckets}), we do not observe a performance drop of the two-tower model on rankings that the logging policy already ranks well. Hence, confounding may not be the correct lens to describe the limitations of the two-tower model. Future research might investigate other avenues, such as conditions for the identifiability of two-tower models as recently described by \citet{Chen2023Identifiability}.

Furthermore, we highlight the striking gap between models trained on clicks and the model trained on expert annotations. The latter vastly outperforms any model trained on clicks on the same dataset by \citet{Supervisor} and \citet{Baidu}, which might point to a discrepancy between user preference and expert annotations on Baidu-ULTR.

Lastly, we wish to address limitations of this research. First, the authors of the dropout method also propose a gradient reversal method, which we did not consider~\cite{DropoutMethod}. Secondly, the backdoor adjustment model was proposed for the DLA method~\cite{Ai2018DLA,UPE}, while we use a classic pointwise two-tower formulation.

\vspace*{-0.5em}
\begin{acks}
This work is supported by the Mercury Machine Learning
Lab, created by TU Delft, the University of Amsterdam, and Booking.com.
\end{acks}

\clearpage
\bibliographystyle{ACM-Reference-Format}
\bibliography{main}

\begin{figure*}[th]
    \centering
    \includegraphics[width=\textwidth]{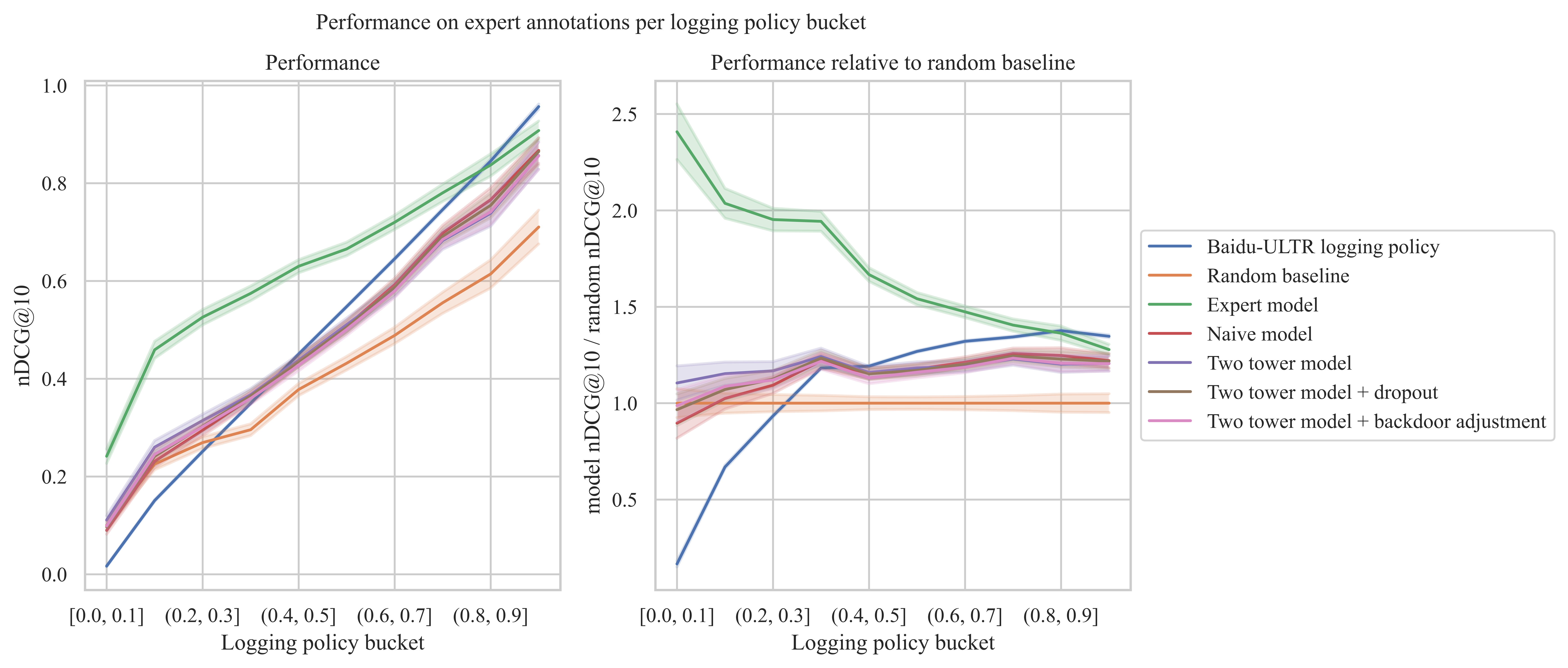}
    \caption{Comparing ranking performance on Baidu-ULTR when test queries are binned by approx. logging policy performance.}
    \label{fig:logging-policy-buckets}
\end{figure*}

\newpage
\appendix

\section{Logging policy estimation}
\label{appendix:logging-policy}
As the Baidu-ULTR dataset does not contain logging policy scores of its production system, we have to estimate the logging policy used to collect the dataset. We do this by training an array of models to predict the logged position of each item from query-document features (to be precise, to predict the reciprocal of each position $1 / k$). Baidu-ULTR contains positions only for documents in the click dataset and not for documents in the test set of expert annotations; therefore, we evaluate how well we can predict positions on a separate test partition of the click dataset. As most rankings contain ten documents, we compare models using nDCG@10. If we perfectly predict the logged position of each document from its features, we would perfectly replicate the Baidu logging policy and achieve an nDCG score of 1.0.

We use the click dataset as provided by \citet{Supervisor} and compared an array of gradient boosting- and neural network-based rankers with different loss functions trained on various amounts of data. We validate and test all models on 25\% of the test click dataset. The best-performing model in our setup is a LightGBM LambdaMART model~\cite{LightGBM} trained on 2\% of the training click dataset ($\approx 290$k query-document pairs) due to computing limitations of our setup. However, the model outperformed all other compared approaches after tuning the number of trees $\in \{300, 500, 1000\}$, the number of leaves $\in \{5, 10, 20\}$, and learning rate $\in \{0.1, 0.01, 0.05\}$. The best-performing LambdaMART model uses 500 trees with 10 leaves and a learning rate of $0.01$.

Lastly, we note that the backdoor adjustment method by \citet{UPE} also requires a logging policy estimation. We follow their paper and use a feed-forward neural network with an attention rank loss~\cite{AttentionLoss}. Otherwise, their training setup is very similar to the one described above, as the prediction target used in their work is also the reciprocal of the logged document position~\cite{UPE}. We use the same validation and testing splits as for the LambdaMART above but train the model on 25\% of the training click dataset. The best neural network-based logging policy estimation for the backdoor adjustment approximates the Baidu logging policy with an nDGC@10 of $0.931\pm4.4\cdot10^{-4}$, which is very close to our LambdaMART model (see Table~\ref{table:rq1}).

\section{Two-tower confounding}
\label{appendix:two-tower-confounding}
We train and validate the two-tower and naive models on 25\% of the pre-processed Baidu-ULTR click dataset by \citet{Supervisor} ($\approx 3.6$m query-document pairs). The final ranking performance is reported on the full expert annotated test set of Baidu-ULTR using nDCG@10 ($\approx 380$k query-document pairs). As relevance input features, all models use BERT-based query-document embeddings and classical learning to rank features (including bm25, tf-idf, and query-likelihood)\footnote{The full list of features is available at:\\\url{https://huggingface.co/datasets/philipphager/baidu-ultr_uva-mlm-ctr}}. Since the values of classical learning to rank features span a large range, we apply $\text{log1p}(x)=\ln(1+|x|)\odot \text{sign}(x)$ scaling as recommended by \citet{Qin2021GBDT}. The bias tower of the two-tower models has only access to the logged item position.

For our neural network-based models (i.e., two-tower and naive models), we tune the number of hidden layers~$\in\{2,3,4\}$ and number of neurons per layer $\in\{64, 128, 256, 512\}$. We find $2$ hidden layers of $256$~neurons and ReLU activations to be optimal in our setup. As the bias towers only use position as an input, we parameterize the bias towers with a single linear layer. Adding more hidden layers did not improve the expressivity of the bias towers. We train for three epochs, use a batch size of 512, and optimize parameters using Adam. We tune the learning rate~$\in\{0.005,0.01,0.03\}$ with the optimal rate being $0.01$ in our setup. Lastly, we select a dropout rate of~$0.3$ for the dropout method after trying out dropout rates~$\in\{0.2,0.3,0.4\}$. We implement our neural networks using Flax~\cite{flax2020github}.

The hyperparameter tuning setup for the LambdaMART model trained directly on expert annotations with 5-fold cross-validation is identical to the LambdaMART model used in RQ1 (Appendix~\ref{appendix:logging-policy}). The expert model is parameterized with $500$ trees with $10$ leaves and a learning rate of $0.01$.

\section{Comparing ranking performance over logging policy performance}
\label{appendix:logging-policy-buckets}
In Figure~\ref{fig:logging-policy-buckets}, we visualize our results from RQ2 when grouping test queries by their approximated logging policy performance. On the left side, we show the absolute model performance, and on the right side, we visualize relative model performance compared to the random baseline. We visualize nDCG performance relative to a random baseline as different test queries have varying amounts of relevant documents, making some queries easier to rank than others. From the theory around the confounding problem, we expect the standard two-tower model to perform worse on queries on which the logging policy already performs well. However, we observe no such trend on the Baidu-ULTR dataset. Nonetheless, we observe that the naive and two-tower models trained on clicks cannot consistently outperform the approximated logging policy. Another interesting observation is that the expert model improves the most over models trained on clicks on queries which the logging policy also struggles with. This might indicate a problem when the logging policy does not expose the most relevant items to the user when gathering click feedback. Lastly, we highlight a limitation of our analysis. Since Baidu-ULTR lacks a parallel corpus of clicked and annotated test queries, we approximated the logging policy using the click dataset. However, if there is a significant distributional shift between clicked and annotated queries, our approximation may not accurately reflect the true logging policy performance on the annotated test set.

\end{document}